\documentclass{article}
\usepackage{graphics}
\title{\bf Torsion-induced spin precession}
\author{Morteza Mohseni\thanks{email:m-mohseni@pnu.ac.ir}
\\{\small Physics Department, Payame Noor University, Tehran 19395-4697,
Iran}}
\begin{document}
\maketitle
\begin{abstract}
\noindent We investigate the motion of a spinning test particle in a spatially-flat FRW-type space-time in the framework of the Einstein-Cartan
theory. The space-time has a torsion arising from a spinning fluid filling the space-time. We show that for spinning particles with nonzero transverse spin components, the torsion induces a precession of particle spin around the direction of the fluid spin. We also show that a charged spinning particle moving in a torsion-less spatially-flat FRW space-time in the presence of a uniform magnetic field undergoes a precession of a different character.

\end{abstract}
\section{Introduction}
Torsion and curvature are important concepts in differential geometry with wide range of applications in physics, particularly in the description
of gravity. To this end, these are used in several different ways. In the general theory of relativity, space-time has curvature but
torsion is absent. On the other hand, in the so-called teleparallel gravity \cite{arc}, space-time has some torsion encoding
the gravity  but no curvature. In the Einstein-Cartan theory both torsion and curvature are present (for a brief review see \cite{tra}, see also
\cite{obu}).

From a gauge theory point of view, the general theory of relativity has been formulated as a gauge theory of Lorentz symmetry \cite{uti} while
the Einstein-Cartan theory can be formulated as gauge theory of Poincar\'e transformations \cite{kib}. In a recent paper \cite{cha} a gauge
theory formulation of noncommutative gravity has been proposed. This formulation is based on the so-called twisted Poincar\'e symmetry and the
general coordinate transformations invariance. Twisted Poincar\'e symmetry and its implications on relativistic invariance in noncommutative space-times has been discussed in \cite{kul,pre}. The underlying symmetry of the noncommutative space-times has been introduced in \cite{kul}
by invoking the twisted Poincar\'e symmetry, a concept originating from the theory of quantum groups. These studies, \cite{cha,kul,pre}, show the importance of torsion for the modern theories of gravitation like noncommutative theories.

Even though the current astronomical data are not able to distinguish between gravitational theories with and without torsion, it could have had important effects in the early universe. In fact it is possible to obtain a metric corresponding to an inflationary universe by taking quantum corrections by vacuum effects to classical gravitational action into account \cite{buch}. The phase structure of the effective potential for the conformal sector of quantum gravity with torsion and its relation to inflationary cosmology has been discussed in \cite{eli}. A spin-dominated inflation has also been proposed in \cite{gas} in the framework of the Einstein-Cartan theory.

The motion of a test particle is not affected by the space-time torsion. From the point of view of particle dynamics the simplest situation in which the torsion comes into the play is the motion of spinning particles. The motion of these particles is governed by the equations first obtained
in \cite{hehl,traut}. Other derivations of the equations of motion of spinning particles may be found in \cite{yas,hoj,cog,cog1,pei,nom,lec}. If one neglects torsion, these equations reduce to the Mathisson-Papapetrou-Dixon (MPD) equations \cite{dix} which describe the motion of a pole-dipole particle in a curved background. However, it should be noted that a pole-dipole rotational spin angular momentum couples to the background curvature but not to the torsion and it is the intrinsic, i.e. elementary particle, spin that couples to torsion \cite{obu,yas}.

The aim of the present work is to study the effect of torsion on the dynamics of a spinning particle in a cosmological model with torsion.
This is mainly motivated by the above considerations and the fact that a natural experimental framework to detect the space-time torsion, if any, is provided by investigating its effects on particle dynamics.

The paper is organized as follows. In the next section we review the Einstein-Cartan theory. Then we consider the motion of particles with an intrinsic spin (treated classically) in a spatially flat FRW-type space-time with torsion and show that the particle spin precess due to a coupling with torsion. Then we investigate the motion of a test body with charge and spin in a spatially flat FRW universe in the presence of a uniform magnetic field and show that it undergoes a spin precession around the magnetic field direction. The last section is devoted to the conclusions.
\section{Cosmological model}
In Einstein-Cartan theory space-time dynamics may be described by the following equations known as Sciama-Kibble equations \cite{kib,sci1,sci}
\begin{equation}
R_{\mu\nu}-\frac{1}{2}Rg_{\mu\nu}=-8\pi t_{\mu\nu}\label{h1}
\end{equation}
and
\begin{equation}
{Q^\rho}_{\mu\nu}-\delta^\rho_\mu{Q^\sigma}_{\nu\sigma}+\delta^\rho_\nu{Q^\sigma}_{\mu\sigma}=8\pi{S^\rho}_{\mu\nu}\label{h2}
\end{equation}
where the first one is the equivalent of the Einstein field equation (in geometric units) and the second one determines the torsion
\begin{equation}
{Q^\rho}_{\mu\nu}=8\pi\left({S^\rho}_{\mu\nu}-\frac{1}{2}\delta^\rho_\mu{S^\sigma}_{\nu\sigma}-\frac{1}{2}\delta^\rho_\nu{S^\sigma}_{\sigma\mu}
\right)\label{h3}.
\end{equation}
Here $t_{\mu\nu}$ is an asymmetric energy-momentum tensor and $S^{\mu\nu\lambda}=-S^{\nu\mu\lambda}$ is the spin density tensor. In the absence of torsion $t_{\mu\nu}$ becomes symmetric as in general relativity. Following a procedure due to Belinfante and Rosenfeld (see e.g. \cite{tra}), it is possible to construct a symmetric energy-momentum tensor $T^{\mu\nu}$ out of $t^{\mu\nu}$ and $S^{\mu\nu\lambda}$. The symmetry of this energy-momentum tensor then leads to an equation of motion for spin (similar to equation (\ref{e876a}) of the next section) \cite{kop}. The energy-momentum tensor
$T^{\mu\nu}$ satisfies the conservation equation $\nabla_\mu T^{\mu\nu}=0.$

In this work we are interested in a particular cosmological solution to the above equations first introduced in \cite{kop}. This space-time has
the following line element
\begin{equation}\label{f9}
ds^2=-dt^2+a^2(t)\delta_{ij}dx^idx^j
\end{equation}
in which ${x^i}, i=1,2,3$ is a set of spatial Cartesian coordinates.
This is a solution of equation (\ref{h1})  with  $$t^{00}=-\rho(t).$$ The torsion is determined from equation (\ref{h2}) with the following non-vanishing components of the spin density tensor $$S_{230}=-S_{320}=S_{23}u_0=-\sigma(t)$$ which in turn satisfy the relevant spin equation of motion. This corresponds to a universe filled with a spinning dust with a spin $S_{23}=\sigma(t)$ in the $x$-direction. The scale factor evolution is determined by a Friedmann-type equation
\begin{equation}\label{scal}
(a^2(t)\dot a(t))^2-2Ma^3(t)+3S^2=0
\end{equation}
in which $M=\frac{4\pi}{3}\rho(t)a^3(t)$ and $S=\frac{4\pi}{3}\sigma(t)a^3(t)$ are constants.
\section{The particle dynamics}
The motion of a spinning particle in a curved background with torsion is governed by a spin evolution equation
\begin{equation}\label{e876a}
 u^\alpha\nabla_\alpha s^{\mu\nu}=p^\mu u^\nu-p^\nu u^\mu
\end{equation}
and a generalized version of the MPD equation \cite{tra}
\begin{equation}\label{e113}
{\dot p}^\mu=g^{\mu\sigma}{Q^\rho}_{\sigma\nu}p_\rho u^\nu-\frac{1}{2}{R^\mu}_{\nu\rho\sigma}u^\nu s^{\rho\sigma}.
\end{equation}
In these equations $p^\mu$ is the particle four-momentum, $s^{\mu\nu}$ its spin tensor, and $u^\mu$ its four-velocity.
According to these equations the four-momentum is not in general in the same direction as the four-velocity.
In the absence of torsion the first term in the right hand side of the above equation disappears, leaving the standard MPD equation.

There are thirteen unknowns (four momentum components, six spin components and three components for the particle trajectories) but
the above equations provide only ten independent equations. The remaining necessary equations are usually obtained by
the so-called supplementary condition. One such condition is given by
\begin{equation}\label{e113a}
p_\mu s^{\mu\nu}=0.
\end{equation}
This is known as Tulczyjew condition and ensures that the particle spin has no electric component in a frame with zero three-velocity.

For space-times possessing some symmetries there are a number of Noether equations giving the constants of motion \cite{hoj2,nie}.
These kind of equations are more useful in situations where the equations of motion are rather complicated. Here we solve the equations of motion
without resort to such Noether equations.
\section{Spin precession}
We are interested in the effect of torsion on the motion of a spinning particle. We consider a spatially-flat FRW type space-time with torsion
described by (\ref{f9}). Nonzero components of the torsion tensor are
\begin{equation}\label{f11}
{Q^3}_{20}=-{Q^2}_{30}=\frac{8\pi\sigma}{a^2}.
\end{equation}
The connection and curvature can be calculated from the following relations \cite{tra}
\begin{eqnarray}
{\Gamma^\mu}_{\lambda\nu}&=&{{\mathring\Gamma}^\mu}_{\lambda\nu}+\frac{1}{2}\left({Q^\mu}_{\lambda\nu}+{{Q_\nu}^\mu}_{\lambda}
+{{Q_\lambda}^\mu}_\nu\right)\label{e875a}\\
{R^\mu}_{\nu\alpha\beta}&=&\partial_\alpha\Gamma^\mu_{\beta\nu}-
\partial_\beta\Gamma^\mu_{\alpha\nu}+\Gamma^\mu_{\alpha\delta}
\Gamma^\delta_{\beta\nu}-\Gamma^\mu_{\beta\delta}\Gamma^\delta_{\alpha\nu}\label{f5}
\end{eqnarray}
where ${{\mathring\Gamma}^\mu}_{\lambda\nu}$ represents the torsion-free part of the connection coefficients.
The non-vanishing components (up to the relevant symmetries) are given by
\begin{eqnarray*}
\Gamma^0_{ii}=a{\dot a},\Gamma^i_{0i}=\frac{{\dot a}}{a},
\Gamma^2_{03}=\frac{4\pi\sigma}{a^2},\Gamma^3_{02}=-\frac{4\pi\sigma}{a^2}
\end{eqnarray*}
and
\begin{eqnarray*}
R_{ijij}&=&(a{\dot a})^2,R_{0i0i}=-a{\ddot a},R_{0203}=-\frac{4\pi\sigma\dot a}{a}
\end{eqnarray*}
respectively. In the above expressions an over-dot means differentiation $\frac{d}{dt}$.

In a co-moving frame, equation (\ref{e113}) reads
\begin{equation}\label{e114}
\frac{dp^\mu}{d\tau}+{\Gamma^\mu}_{0\beta}p^\beta={{Q_\rho}^\mu}_0p^\rho-\frac{1}{2}{R^\mu}_{0\rho\sigma}s^{\rho\sigma}
\end{equation}
and thus
\begin{eqnarray}
\frac{dp^0}{d\tau}&=&0,\label{g1}\\
\frac{dp^1}{d\tau}+\frac{\dot a}{a}p^1&=&-\frac{\ddot a}{a}s^{01},\label{g2}\\
\frac{dp^2}{d\tau}+\frac{\dot a}{a}p^2-\frac{4\pi\sigma}{a^2}p^3&=&-\frac{\ddot a}{a}s^{02}-\frac{4\pi\sigma\dot a}{a^3}s^{03},\label{g3}\\
\frac{dp^3}{d\tau}+\frac{\dot a}{a}p^3+\frac{4\pi\sigma}{a^2}p^2&=&-\frac{\ddot a}{a}s^{03}-\frac{4\pi\sigma\dot a}{a^3}s^{02}.\label{g4}
\end{eqnarray}
The spin equation (\ref{e876a}) reduces to
\begin{eqnarray}
\frac{ds^{01}}{d\tau}+\frac{\dot a}{a}s^{01}&=&-p^1,\label{g9}\\
\frac{ds^{02}}{d\tau}+\frac{\dot a}{a}s^{02}+\frac{4\pi\sigma}{a^2}s^{03}&=&-p^2,\label{g10}\\
\frac{ds^{03}}{d\tau}+\frac{\dot a}{a}s^{03}-\frac{4\pi\sigma}{a^2}s^{02}&=&-p^3,\label{g11}\\
\frac{ds^{12}}{d\tau}+2\frac{\dot a}{a}s^{12}+\frac{4\pi\sigma}{a^2}s^{13}&=&0,\label{g12}\\
\frac{ds^{13}}{d\tau}+2\frac{\dot a}{a}s^{13}-\frac{4\pi\sigma}{a^2}s^{12}&=&0,\label{g13}\\
\frac{ds^{23}}{d\tau}+2\frac{\dot a}{a}s^{23}&=&0\label{g14}.
\end{eqnarray}
The simplest solution to the above systems of equations compatible with condition (\ref{e113a}) is as follows
\begin{eqnarray}
p^0&=&m,\label{g1a}\\
p^i&=&0,\label{g1b}\\
s^{0i}&=&0,\label{g1c}\\
s^{23}(\tau)&=&\sigma^La^{-2}(\tau),\label{g14a}\\
s^{12}(\tau)&=&\sigma^Ta^{-2}(\tau)\cos\left(4\pi\int\frac{\sigma}{a^2}d\tau\right),\label{g12a}\\
s^{13}(\tau)&=&\sigma^Ta^{-2}(\tau)\sin\left(4\pi\int\frac{\sigma}{a^2}d\tau\right)\label{g13a}
\end{eqnarray}
with $m$ being a constant which can be taken as the mass of the particle and $\sigma^L,\sigma^T$ are two constants determined by the particle initial spin orientation. The first three equations of the above set mean that in this case the particle four-momentum and four-velocity are co-linear. The last three equations explicitly show that the particle spin precess around an axis along the direction of spin of the fluid filling the universe provided the particle has a non-vanishing initial transverse component. The characteristics of this precession are controlled by torsion ${Q^\mu}_{\nu\lambda}$. In the absence of torsion the above equations describe a particle with its momentum and velocity being co-linear and the evolution of its spin depending only on the evolution of the universe scale factor.

\begin{figure}
\resizebox{\textwidth}{!}
{\includegraphics{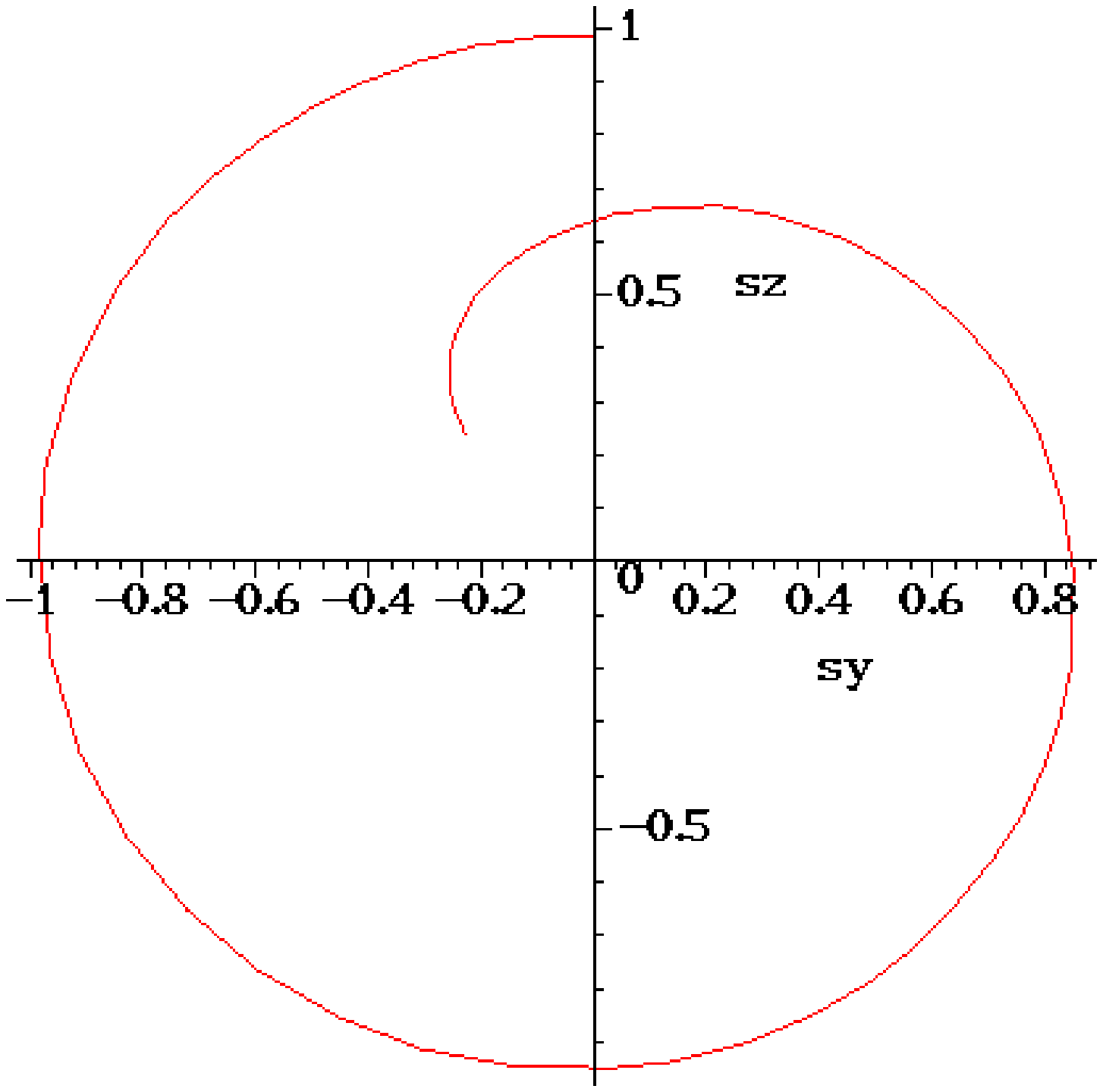}\includegraphics{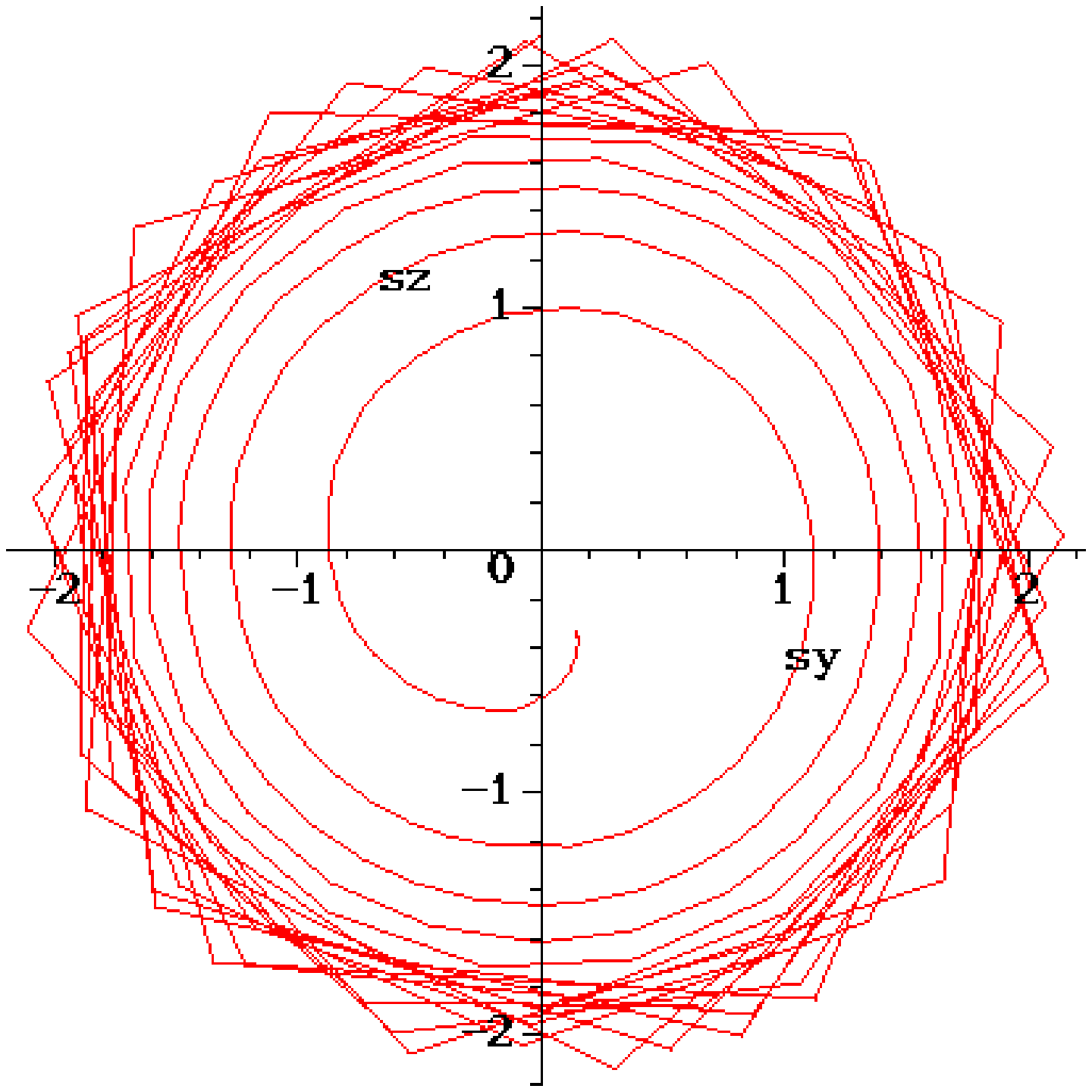}}
\caption{The path of spin vector $s^\mu$ in the transverse plane in units of $\sigma^T\beta$. The values of the parameters are fixed by setting
$\alpha=1$ (left), and $\alpha=0.1$ (right). For both graphs $\beta=1$ and $0\leq t\leq5$. }
\label{fig:1}
\end{figure}

The particle spin can also be described by a spin density four-vector
\begin{equation}
s^\mu=\frac{1}{2\sqrt{-g}}\varepsilon^{\mu\nu\alpha\beta}p_\nu s_{\alpha\beta}
\end{equation}
in which $\varepsilon^{\mu\nu\alpha\beta}$ is the alternating symbol. Thus by choosing $\varepsilon^{0123}=+1$ we have
\begin{eqnarray}
s^1(\tau)&=&\sigma^{L}a^{-1}(\tau),\label{g14l}\\
s^2(\tau)&=&-\sigma^{T}a^{-1}(\tau)\sin\left(4\pi\int\frac{\sigma}{a^2}d\tau\right),\label{g12l}\\
s^3(\tau)&=&\sigma^{T}a^{-1}(\tau)\cos\left(4\pi\int\frac{\sigma}{a^2}d\tau\right)\label{g13l}.
\end{eqnarray}
For
\begin{equation}\label{eu}
a(t)=\sqrt{\frac{3}{2M}}\left(t^2+\frac{S^2}{3M^2}\right)^{\frac{1}{3}}
\end{equation}
which is a non-singular solution to (\ref{scal}) (as long as the shear is absent, see also \cite{ray}), and noting that $$4\pi\sigma(t)=3Sa^{-3}(t),$$ we obtain
\begin{eqnarray}
s^2(\tau)&=&\frac{-\sigma^T\beta}{(t^2+\alpha^2)^{\frac{1}{3}}}\sin\left(\gamma t{}_2F_1\left(\frac{1}{2},\frac{5}{3}
;\frac{3}{2};-\frac{t^2}{\alpha^2}\right)\right),\label{g12h}\\
s^3(\tau)&=&\frac{\sigma^T\beta}{(t^2+\alpha^2)^{\frac{1}{3}}}\cos\left(\gamma t{}_2F_1\left(\frac{1}{2},\frac{5}{3}
;\frac{3}{2};-\frac{t^2}{\alpha^2}\right)\right)\label{g13h}
\end{eqnarray}
in which $\alpha=\frac{S}{\sqrt{3}M}$, $\beta=\sqrt{\frac{2M}{3}}$, $\gamma=\frac{9\sqrt{3}}{2}\alpha^{-\frac{7}{3}}\beta^7$, and $${}_2F_1(a,b;c;x)$$ is the hypergeometric function. This spin precession is depicted in figure \ref{fig:1} which shows high sensitivity with respect to torsion.

To gain a better understanding of this nontrivial behavior, it would be helpful to compare it with a more familiar situation, namely spin precession due to an interaction with a magnetic field in general relativity which is studied in the subsequent section.
\section{Precession induced by a magnetic field}
\begin{figure}[h]
\resizebox{\textwidth}{!}
{\includegraphics{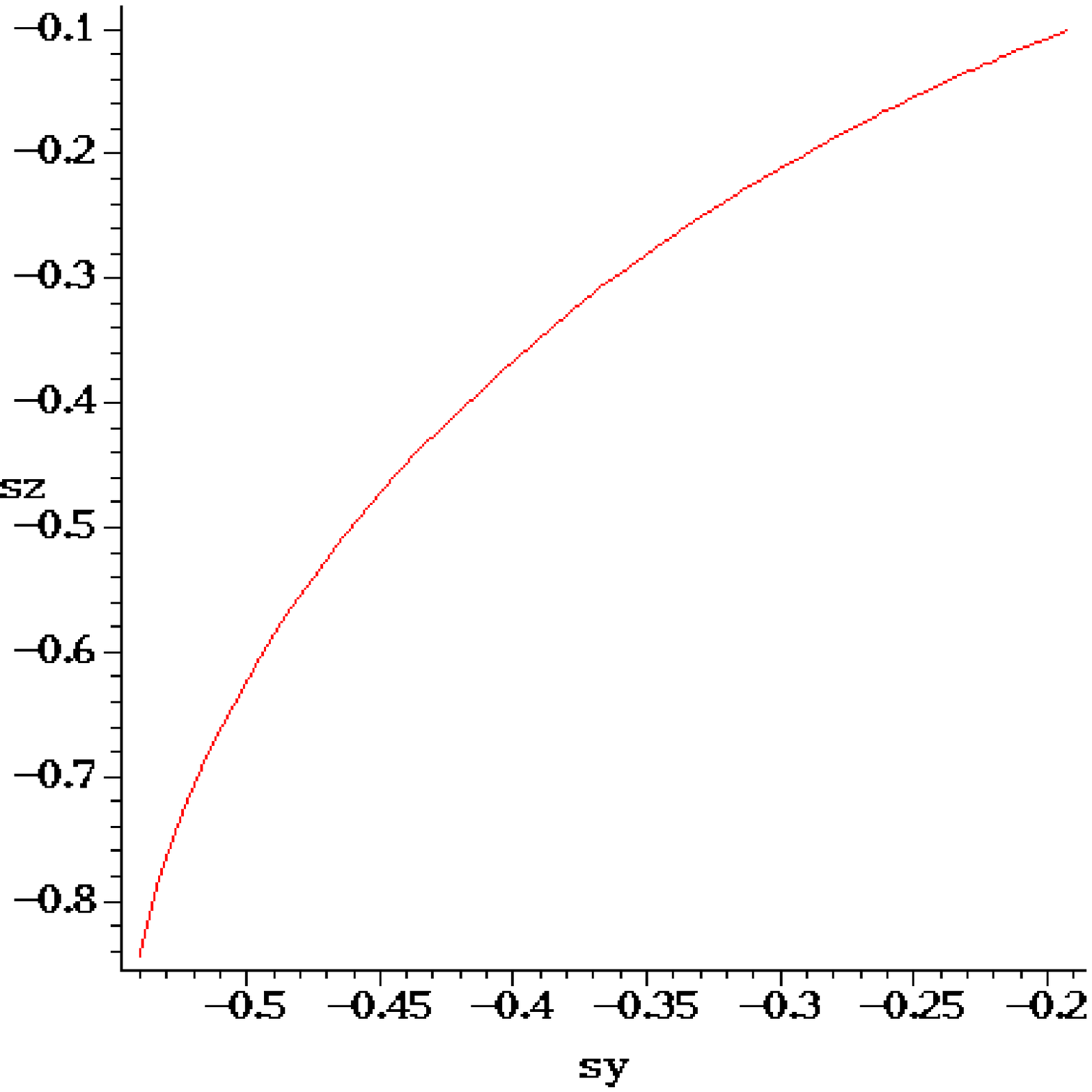}\includegraphics{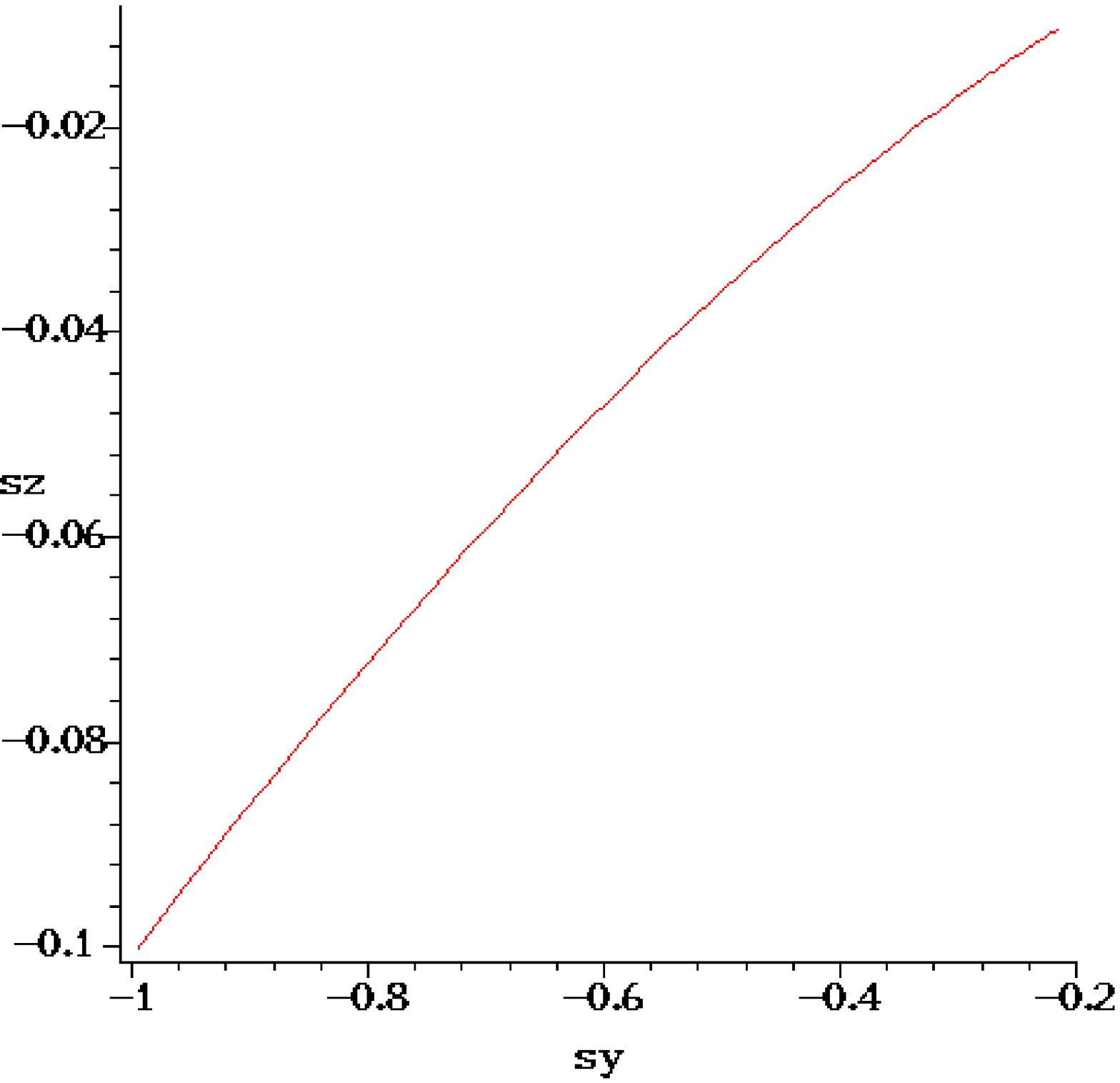}}
\caption{The path of spin vector $s^\mu$ in the transverse plane in units of $\sigma^T\beta$ for $\beta=1$, $Bk=\frac{1}{3}$ (left), $Bk=\frac{1}{30}$ (right). For both graphs $1\leq t\leq 10$.}
\label{fig4}
\end{figure}

In this section we consider a charged spinning particle moving in a spatially flat FRW space-time (no torsion) in the presence of a uniform magnetic field.
One can use the Dixon-Souriau (DS) equations \cite{dix2,sou} to describe this motion. We have
\begin{eqnarray}
{\dot s}^{\mu\nu}&=&p^\mu u^\nu-p^\nu u^\mu-k(s^{\mu\kappa}{F_\kappa}^\nu-s^{\nu\kappa}{F_\kappa}^\mu),\label{j1}\\
{\dot p}^\mu&=&-\frac{1}{2}{R^\mu}_{\nu\lambda\rho}s^{\lambda\rho}u^\nu+q{F^\mu}_\beta u^\beta+\frac{k}{2}s^{\kappa\rho}\nabla^\mu F_{\kappa\rho}
\label{j2}
\end{eqnarray}
in which $k=\frac{qg}{2m}$ is constant, $q$ is the particle electric charge, $g$ is the particle gyromagnetic ratio, and $F_{\mu\nu}$ represents the electromagnetic tensor. The condition (\ref{e113a}) is still assumed. The space-time metric is given by equation (\ref{f9}). We consider an electromagnetic field represented by
\begin{equation}\label{s1}
F^{23}=\frac{B}{a^4}
\end{equation}
which is a solution to the source-free Maxwell equations
\begin{equation}\label{max1}
\nabla_\nu F^{\mu\nu}=0,
\end{equation}
\begin{equation}\label{max2}
\nabla_\lambda F_{\mu\nu}+\nabla_\mu F_{\nu\lambda}+\nabla_\nu F_{\lambda\mu}=0
\end{equation}
in a background spatially-flat FRW space-time. This corresponds to
\begin{eqnarray*}
B^\mu&=&\frac{1}{2\sqrt{-g}}\varepsilon^{\mu\nu\alpha\beta}u_\nu F_{\alpha\beta}\\&=&\frac{B}{a^3}\varepsilon^{\mu 023}
\end{eqnarray*}
a uniform magnetic field in the $x$-direction.
By inserting this in the DS equations we will reach at equations similar to (\ref{g1})-(\ref{g14}) with the following solution
\begin{eqnarray}
p^0&=&m,\label{w1a}\\
p^i&=&0,\label{w1b}\\
s^{0i}&=&0,\label{w1c}\\
s^{23}(\tau)&=&\sigma^La^{-2}(\tau),\label{w14a}\\
s^{12}(\tau)&=&\sigma^Ta^{-2}(\tau)\sin\left(Bk\int\frac{1}{a^2}d\tau\right),\label{w12a}\\
s^{13}(\tau)&=&\sigma^Ta^{-2}(\tau)\cos\left(Bk\int\frac{1}{a^2}d\tau\right)\label{w13a}
\end{eqnarray}
which for a nonzero $\sigma^T$ shows a spin precession around the direction of the magnetic field in the following form
\begin{eqnarray}
s^2(\tau)&=&-\sigma^Ta^{-1}(\tau)\cos\left(Bk\int\frac{1}{a^2}d\tau\right),\label{w12f}\\
s^3(\tau)&=&\sigma^Ta^{-1}(\tau)\sin\left(Bk\int\frac{1}{a^2}d\tau\right)\label{w13f}.
\end{eqnarray}
This is depicted in figure \ref{fig4} for a scale factor of the form (\ref{eu}) but with $S=0$.
\section{Conclusions}
We have solved the equations of motion of a test particle with spin in flat FRW-type space-time with torsion in the framework
of the Einstein-Cartan theory. The space-time torsion originates from a spinning fluid filling the space with one non-vanishing spin component.
By choosing a co-moving frame we focused on the evolution of the particle spin. The equations of motion admit a solution that for a non-vanishing initial transverse spin component explicitly represents a spin precession due to coupling of particle spin with the space-time torsion, or equivalently, with the spinning fluid spin. This precession is around an axis along the direction of the fluid spin and its characteristics depend on the ratio of fluid square spin density to its energy density. For an initial longitudinal initial spin the spin orientation does not change as the scale factor evolves.

We  showed that a charged particle with spin in a FRW space-time with a uniform magnetic field undergoes a precession around the direction of the magnetic field. In fact in these models both torsion and the uniform magnetic field make a space direction a preferred one causing a spin precession around it but with quite different patterns. Also there is another important distinction between these two situations, in the first one (torsion case) the spin can only be an elementary particle spin whereas in the second one (torsion-less case) it could be a rotational spin. These results might be of interest to the experimental research on torsion and the related problem of effective local Lorentz invariance violation \cite{prl} and also to
inflationary cosmology.

The solution presented here corresponds to a situation in which the particle four-momentum and four-velocity are in the same direction. However other types of solutions for which this co-linearity does not hold may also be found by relaxing the supplementary equation (\ref{e113a}).

\section*{Acknowledgements}
I would like to thank Juan A. Nieto for valuable comments.

\end{document}